\documentclass[11pt,a4paper]{article}
\pdfoutput=1
\usepackage{jheppub}
\usepackage[utf8]{inputenc}

\usepackage{color}
\usepackage{amsmath}
\usepackage{pifont}
\usepackage{bbold}

\usepackage{bbm}
\usepackage{verbatim}   
\usepackage{subfigure}
\usepackage{acronym}

\usepackage{amsfonts}
\usepackage{amssymb}
\usepackage{mathrsfs}
\usepackage{graphicx}
\usepackage{multirow}
 \usepackage{slashed}

\usepackage{soul}

\usepackage{caption}
\usepackage{tikz}
\usetikzlibrary{decorations.pathreplacing}

\usepackage{subfiles}

\title{The Photophobic ALP}

\author[a,c]{Nathaniel Craig,}
\emailAdd{ncraig@physics.ucsb.edu}
\author[b,c]{Anson Hook}
\emailAdd{hook@umd.edu}
\author[a]{and Skyler Kasko}
\emailAdd{skasko@physics.ucsb.edu}

\affiliation[a]{Department of Physics, University of California, Santa Barbara, CA 93106, USA}
\affiliation[b]{Maryland Center for Fundamental Physics, Department of Physics, University of Maryland, College Park, MD 20742, USA}
\affiliation[c]{Kavli Institute for Theoretical Physics, Santa Barbara, CA 93106, USA}

\abstract{We explore models and phenomenology of a photophobic axion-like particle (ALP), an axion whose coupling to photons is maximally suppressed without fine-tuning of the underlying parameters. We demonstrate that photophobia can be a natural UV property of ALP models and determine the irreducible coupling of photophobic ALPs to photons induced by violations of the axion shift symmetry.
The signatures of photophobic axions are radically different from those of typical ALPs and are of particular interest for collider-based experiments, for which Standard Model triboson measurements provide a significant probe at higher masses. A variety of terrestrial and astrophysical measurements constrain the parameter space of photophobic ALPs, though bounds are typically much weaker compared to typical ALPs.  We discuss implications for particle production relaxion models based on photophobic ALPs, finding that they are in mild tension with existing experimental constraints.}

\begin{document} 
\maketitle
\flushbottom

\section{Introduction}
\label{sec:intro}

Light pseudoscalars are motivated candidates for physics beyond the Standard Model (SM) and novel targets for current and future experimental searches. The most famous among these is the QCD axion, which may both resolve the strong CP problem~\cite{Peccei:1977hh,Peccei:1977ur,Weinberg:1977ma,Wilczek:1977pj} and account for dark matter~\cite{Preskill:1982cy, Abbott:1982af, Dine:1982ah}. But the QCD axion is only one example of a much broader class of axion-like particles (ALPs) that may solve problems of the Standard Model~\cite{Graham:2015cka} or simply point to the existence of additional sectors beyond our own. 

Such ALPs can enjoy a variety of couplings to the Standard Model consistent with a shift symmetry and underlying CP invariance, leading to a range of possible experimental signatures (e.g.~\cite{Masso:1995tw, Rupak:1995kg, Raffelt:1996wa, Asztalos:2009yp,Irastorza:2013dav, Mimasu:2014nea, Alekhin:2015byh, Jaeckel:2015jla, Anastassopoulos:2017ftl, Brivio:2017ije}). Thus far, significant attention has been focused on the coupling of ALPs to pairs of photons, as ALP-photon couplings are generic and provide the dominant decay mode for sufficiently light ALPs. This coupling leads to a distinctive and thoroughly-explored ALP phenomenology. Recent studies (e.g.~\cite{Izaguirre:2016dfi,Brivio:2017ije,Bauer:2017ris, Dolan:2017osp}) have broadened to include the full set of ALP couplings to electroweak gauge bosons implied by gauge invariance, which can furnish distinctive associated production modes as well as additional loop-induced couplings. But only limited attention has been given to so-called {\it photophobic ALPs}, whose leading couplings to Standard Model electroweak gauge bosons do not involve a pair of photons.\footnote{See~\cite{Brivio:2017ije} for some collider constraints on photophobic ALPs assuming the axion is collider-stable. It bears emphasizing that {\it hadronic} axions with leading couplings to gluon pairs are typically far from photophobic, as they inherit a large coupling to photon pairs via mixing with pseudoscalar mesons.  It is possible to cancel a tree level coupling against this induced coupling~\cite{Kaplan:1985dv,DiLuzio:2016sbl,DiLuzio:2017pfr}, but it requires ``tuning" in model space.} Of course, such ALPs are not truly photophobic, in that their leading couplings to electroweak gauge bosons give rise to mass-suppressed ALP-photon couplings at one and two loops, but this additional loop suppression is sufficient to significantly alter the phenomenology of light ALPs.  Such photophobic ALPs are typically considered non-generic (e.g.~\cite{Bauer:2017ris}), since photophobia requires a cancellation between ostensibly unrelated couplings to $SU(2)_L$ and $U(1)_Y$ gauge bosons.  Even so, their novel phenomenology and potential relevance to the electroweak hierarchy problem via the particle production relaxation mechanusm~\cite{Hook:2016mqo} make them an attractive target for further study. 

In this paper, we place photophobia on firmer footing by demonstrating the generic conditions under which a light ALP can avoid leading couplings to the photon {\it without} fine tuning. Of course, such photophobia is only a property of the UV theory, as photophobic couplings to electroweak gauge bosons lead to ALP-fermion couplings at one loop and ALP-photon couplings at both one and two loops \cite{Bauer:2017ris}. Having justified photophobic boundary conditions in the UV, we study the full set of IR couplings inherited by photophobic ALPs; their phenomenological consequences; and existing experimental constraints. In general, the bounds on photophobic ALPs are considerably weaker than their photophilic relatives, opening a wide range of parameter space to potential future experimental tests. 

The paper is organized as follows: In Section \ref{sec:spurion} we present a general spurion argument demonstrating that it is possible for axion-like particles to possess leading couplings to electroweak gauge bosons that do not involve pairs of photons without fine-tuning.  Additional couplings are generated by Standard Model running to the IR, which we compute explicitly.  In Section \ref{sec:pheno} we turn to the phenomenology of photophobic ALPs, including both the dominant production rates and all relevant two- and three-body decay modes into Standard Model final states. We turn to astrophysical and collider limits on photophobic ALPs in Section \ref{sec:limits}, determining the allowed parameter space as a function of the photophobic ALP mass and UV couplings. Although our discussion of photophobic ALP phenomenology is entirely general, in Section \ref{sec:relaxion} we focus on its implications for the particle-production relaxion, a relaxion model that utilizes a  photophobic ALP. 
We conclude in Section \ref{sec:conclusion}. We reserve an explicit photophobic ALP model for Appendix \ref{app:models} and a summary of experimental interpretations for Appendix \ref{app:limitapp}.

\section{Photophobic ALPs}
\label{sec:spurion}

\newcommand\eea{\end{eqnarray}}
\newcommand\bea{\begin{eqnarray}}

It is common lore that photophobic ALPs are tuned, in the sense that an ALP coupling to $SU(2)_L$ and $U(1)_Y$ gauge bosons will generically couple to pairs of photons in the broken symmetry phase.  In this section, we use a spurion argument to demonstrate to what extent a light ALP can avoid coupling to pairs of photons without fine tuning.  An explicit model that realizes the spurion analysis is given in Appendix~\ref{app:models}.  

An ALP can have two types of couplings.  The first type are couplings to vector field strengths and their duals, namely
\bea
\nonumber
\mathcal{L} \supset g_s^2 \left(\frac{a}{4 f_{GG}} + \frac{\theta_3}{32 \pi^2} \right) G \tilde G + g^2 \left (\frac{a}{4 f_{WW}} + \frac{\theta_2}{32 \pi^2} \right ) W \tilde W + g'^2 \left (\frac{a}{4 f_{BB}} +  \frac{\theta_1}{32 \pi^2} \right) B \tilde B 
\eea
where $\theta_{1,2,3}$ are the theta angles for the $U(1)$, $SU(2)$ and $SU(3)$ respectively.  The second type are couplings to fermions of the form
\bea
\mathcal{L} \supset \frac{\partial_\mu a}{f_F} \bar \psi \gamma^\mu \gamma^5 \psi
\eea

How axions couple and how these couplings are renormalized can be understood from the statement that the fermion couplings are shift symmetric while the gauge boson couplings are not, and instead have a topological shift symmetry-breaking spurion $\theta$.  The photophobic ALP corresponds to the case where $1/f_{GG} = 0$ and $f_{WW} = - f_{BB} = f_a$ in the UV, for which the ALP does not couple to pairs of photons.  This UV boundary condition is itself not a tuning, as it can arise naturally in e.g.~left-right symmetric models; we present an explicit model realizing this boundary condition in Appendix \ref{app:models}. In this explicit model, $1/f_F$ is also zero in the UV.

Provided these boundary conditions in the UV, it is then a question of the extent to which they are preserved by RG evolution to the IR.   In the photophobic case of $1/f_{GG} = 0$ and $f_{WW} = - f_{BB} = f_a$, the axion has a spurious shift symmetry 
\bea
\label{Eq: symmetry}
a \rightarrow a + \epsilon f_a \qquad \theta_2 \rightarrow \theta_2 - 8 \pi^2 \epsilon \qquad \theta_1 \rightarrow \theta_1 + 8 \pi^2 \epsilon
\eea
This spurious shift symmetry controls the RG of the axion very tightly.

Let us first consider the case of a massless axion.  
Because $\theta_1$ and $\theta_2$ are multiplying topological quantities, they never appear in perturbation theory.  Thus as RG evolution occurs, $\theta_1$ and $\theta_2$ never appear in any other operators.  Because these parameters dictate any shift symmetry breaking couplings that the axion can have, the axion has $1/f_{GG} = 0$ and $f_{WW} = - f_{BB} = f_a$ not only in the UV but also in the IR.  Thus, since the coupling to the photon has $1/f_{\gamma \gamma} = 1/f_{WW} + 1/f_{BB} = 0$ the axion never couples to the photon!  RG evolution cannot violate symmetries and thus the ALP-photon coupling $1/f_{\gamma \gamma} = 0$ for the massless axion to all orders in perturbation theory.

This strong statement holds only for the axion gauge boson couplings.  The axion fermion couplings are shift symmetric and because the derivative acts on the axion, obey the shift symmetry in Eq.~\ref{Eq: symmetry}.  Thus because they obey all symmetries, it is not surprising to see that they are in fact generated by 1-loop RG evolution. Indeed, 1-loop RG evolution of the boundary conditions $f_{WW} = - f_{BB} = f_a$ at the UV scale $\Lambda$ down to the weak scale generates the coupling to fermions~\cite{Bauer:2017ris}
\bea
\label{Eq: fermions}
\frac{1}{f_F} = - \frac{3 \alpha^2}{4 f_a}\left [ \frac{3}{4 \sin^4 \theta_w} - \frac{1}{\cos^4 \theta_w} ( Y_{F_L}^2 + Y_{F_R}^2 ) \right ] \log \frac{\Lambda^2}{m_W^2}
\eea
where $Y_{F_{L,R}}$ are the hypercharges of the left and right handed fields of the fermion $F$. As long as the axion remains massless, radiative corrections from these ALP-fermion couplings do not themselves generate ALP-photon couplings.

The previous discussion changes when the axion obtains a mass, as a mass term explicitly breaks the shift symmetry shown in Eq.~\ref{Eq: symmetry} and is the spurion of shift symmetry breaking.  Thus the previous statements only hold up to corrections due to the axion mass. Indeed, mass-proportional ALP-photon couplings are generated from the photophobic ALP-vector coupling $1/f_a$ and ALP-fermion couplings $1/f_F$ at one loop. The one loop contributions to the axion photon coupling are \cite{Bauer:2017ris}
\bea \label{Eq: photons}
\frac{1}{f_{\gamma \gamma}} = \frac{2 \alpha}{\pi \sin^2 \theta_w f_a} B_2 \left (\frac{4 m_W^2}{m_a^2} \right) + \sum_F \frac{N_c^F Q_F^2}{2 \pi^2 f_F} B_1\left(\frac{4 m_F^2}{m_a^2}\right)
\eea
where $Q_F$ and $N_c^F$ are the electric charge and color multiplicity for the fermion respectively.  The fact that the spurious symmetry is restored in the limit $m_a \rightarrow 0$ means that the functions $B_{1,2} \propto m_a^2$ as $m_a$ approaches zero.\footnote{$m_a^2$ is the spurion of shift symmetry breaking, not $m_a$, which is why $B_{1,2}$ go to zero quadratically as $m_a^2 \rightarrow 0$.} This limiting behavior can be seen explicitly from the form of the functions $B_{1,2}$ \cite{Bauer:2017ris}
\bea
\begin{aligned}
    B_1(x) &= 1 - x f(x)^2\\
    B_2(x) &= 1 - (x - 1) f(x)^2
  \end{aligned}
 \qquad  \qquad f(x)=\begin{cases}
    \text{arcsin} \frac{1}{\sqrt{x}} & x \geq 1\\
    \frac{\pi}{2} + \frac{i}{2} \log \frac{1 + \sqrt{1-x}}{1 - \sqrt{1-x}} & x < 1.
  \end{cases}
\eea
As required by the spurious symmetry, these functions behave as $B_1 \rightarrow - m_a^2/12 m_F^2$ and $B_2 \rightarrow m_a^2/6 m_W^2$ as $m_a \rightarrow 0$.

Thus we see that the coupling to the photon cannot be completely suppressed, as the shift symmetry preventing the coupling to the photon is broken. Starting from the maximally photophobic boundary conditions $f_{WW} = - f_{BB} = f_a$ and $f_F = 0$, a massive ALP will generically couple to pairs of photons at one loop and $\mathcal{O}(m_a^2/m_W^2)$, and at two loops and $\mathcal{O}(m_a^2/m_F^2)$, as first observed in \cite{Bauer:2017ris}. The latter, two-loop coupling is most relevant for light ALPs at or below the mass scale of light Standard Model fermions. In what follows, we will take the boundary $f_{WW} = - f_{BB} = f_a$ and $f_F = 0$ at the UV scale $\Lambda$ to define the photophobic ALP, with natural ALP-fermion and ALP-photon couplings arising at one and two loops. This is the most ``photophobic'' an ALP can be without genuine fine-tuning of couplings.\footnote{Of course, it is possible to choose $f_{WW}, f_{BB}$ and non-universal $f_F$ at the UV cutoff $\Lambda$ so that $f_{\gamma \gamma} = 0$ at some specific scale. But this results from a genuine tuning and is not stable under RG evolution.}

\section{Photophobic Phenomenology}
\label{sec:pheno}

We now turn to the phenomenology of the photophobic ALP. Photophobic boundary conditions lead to a distinctive pattern of couplings quite unlike those of a generic ALP. Provided the boundary conditions $f_{WW} = - f_{BB} \equiv f_a$ and $f_F = 0$ at the UV scale $\Lambda$, an ALP at or below the weak scale will enjoy the following hierarchy of couplings to Standard Model particles: 
\begin{itemize}
\item Tree-level $\mathcal{O}\left( \frac{1}{f_a} \right)$ couplings to pairs of electroweak gauge bosons, specifically $ZZ, Z\gamma,$ and $WW$.
\item One-loop $\mathcal{O}\left( \frac{1}{16 \pi^2} \frac{1}{f_a} \log(\Lambda^2/m_W^2) \right)$ couplings to pairs of Standard Model fermions lighter than the ALP (with additional $m_a^2/m_F^2$ suppression for $m_a < m_F$) and $\mathcal{O}\left( \frac{1}{16 \pi^2} \frac{m_a^2}{m_W^2} \frac{1}{f_a} \right)$ couplings to $\gamma \gamma$.
\item Two-loop $\mathcal{O} \left( \frac{1}{(16 \pi^2)^2} \frac{m_a^2}{m_F^2} \frac{1}{f_a}  \log(\Lambda^2/m_W^2)  \right)$ couplings to $\gamma \gamma$.
\end{itemize}  
Each of these couplings can be significant for the phenomenology of the photophobic ALP, with relevance varying depending on the ALP mass. Note that couplings to gluons are also generated at two loops (rendering the photophobic ALP more gluophobic than anything else!), but will be largely neglected in what follows as subdominant to the one-loop quark couplings. 

\subsection{ALP Decays}

The distinctive branching ratios of the photophobic ALP are governed by a combination of the intrinsic couplings discussed above; chirality suppression; and phase space considerations. The partial widths for two-body decays into Standard Model particles are, when kinematically open,
\begin{eqnarray}
\Gamma(a \rightarrow Z Z) &=& \frac{1}{32 \pi} \frac{4 e^4 \sin^2 (4 \theta_W)}{\sin^6 (2 \theta_W)} \frac{m_a^3}{f_a^2} \left(1 - \frac{4 m_Z^2}{m_a^2} \right)^{3/2}\\
\Gamma(a \rightarrow Z \gamma) &=& \frac{1}{32 \pi} \frac{e^4}{\sin^2 \theta_W \cos^2 \theta_W} \frac{m_a^3}{f_a^2} \left( 1 - \frac{m_Z^2}{m_a^2} \right)^3 \\
\Gamma(a \rightarrow WW) &=& \frac{1}{32 \pi} \frac{e^4}{\sin^4 \theta_W} \frac{m_a^3}{f_a^2} \left( 1 - \frac{4 m_W^2}{m_a^2} \right)^{3/2}\\
\Gamma(a \rightarrow \gamma \gamma) &=& \frac{1}{64 \pi} e^4 \frac{m_a^3}{f_{\gamma \gamma}^2} \\
\Gamma(a \rightarrow \ell \bar \ell ) &=& \frac{1}{2 \pi} \frac{m_a m_\ell^2}{f_\ell^2} \left( 1 - \frac{4 m_\ell^2}{m_a^2} \right)^{1/2} \\
\Gamma(a \rightarrow Q \bar Q) &=& \frac{3}{2 \pi} \frac{m_a m_Q^2}{f_Q^2} \left(1 - \frac{4 m_Q^2}{m_a^2} \right)^{1/2} \\
\Gamma(a \rightarrow {\rm hadrons}) &=& \frac{1}{8 \pi^3} \alpha_s^2 m_a^3 \left( 1 + \frac{83}{4} \frac{\alpha_s}{\pi} \right) \left| \sum_{q = u,d,s} \frac{1}{f_q} \right|^2
\end{eqnarray}
where the various $f_F$ (with $F= \ell, q, Q$ where $q = u,d,s$ and $Q = c,b,t$) and $f_{\gamma \gamma}$ are given in Eq.s \ref{Eq: fermions} and \ref{Eq: photons}, respectively. Here the inclusive decay rate into hadrons \cite{Bauer:2017ris} neglects light quark masses and assumes $m_a \gg \Lambda_{QCD}$ (for a detailed treatment of hadronic decays around $m_a \sim \Lambda_{QCD}$, see \cite{Bauer:2017ris}).  In addition to these two-body decays, three-body decays via $V V^*$ are numerically relevant for $m_a \lesssim m_Z$.

For sufficiently heavy ALPs, decays are dominated by tree-level couplings to vector bosons. In practice, these two- and three-body vector boson decay modes dominate for $m_a \gtrsim 50$ GeV, at which point loop-induced couplings become competitive with the phase space suppression of the three-body decays. For lighter ALPs, the loop-induced couplings to fermions provide the dominant decay channels down to $m_a = 2 m_e$, below which the suppressed coupling to photons provide the leading decay channel. This differs significantly from generic ALPs, as the $a \rightarrow \gamma \gamma$ mode is only significant once the fermionic decay modes (neutrinos excepted) are closed. The partial decay widths as a function of the ALP mass are illustrated in Figure~\ref{fig:gamma}.

\begin{figure}[t] 
   \centering
   \includegraphics[width=4in]{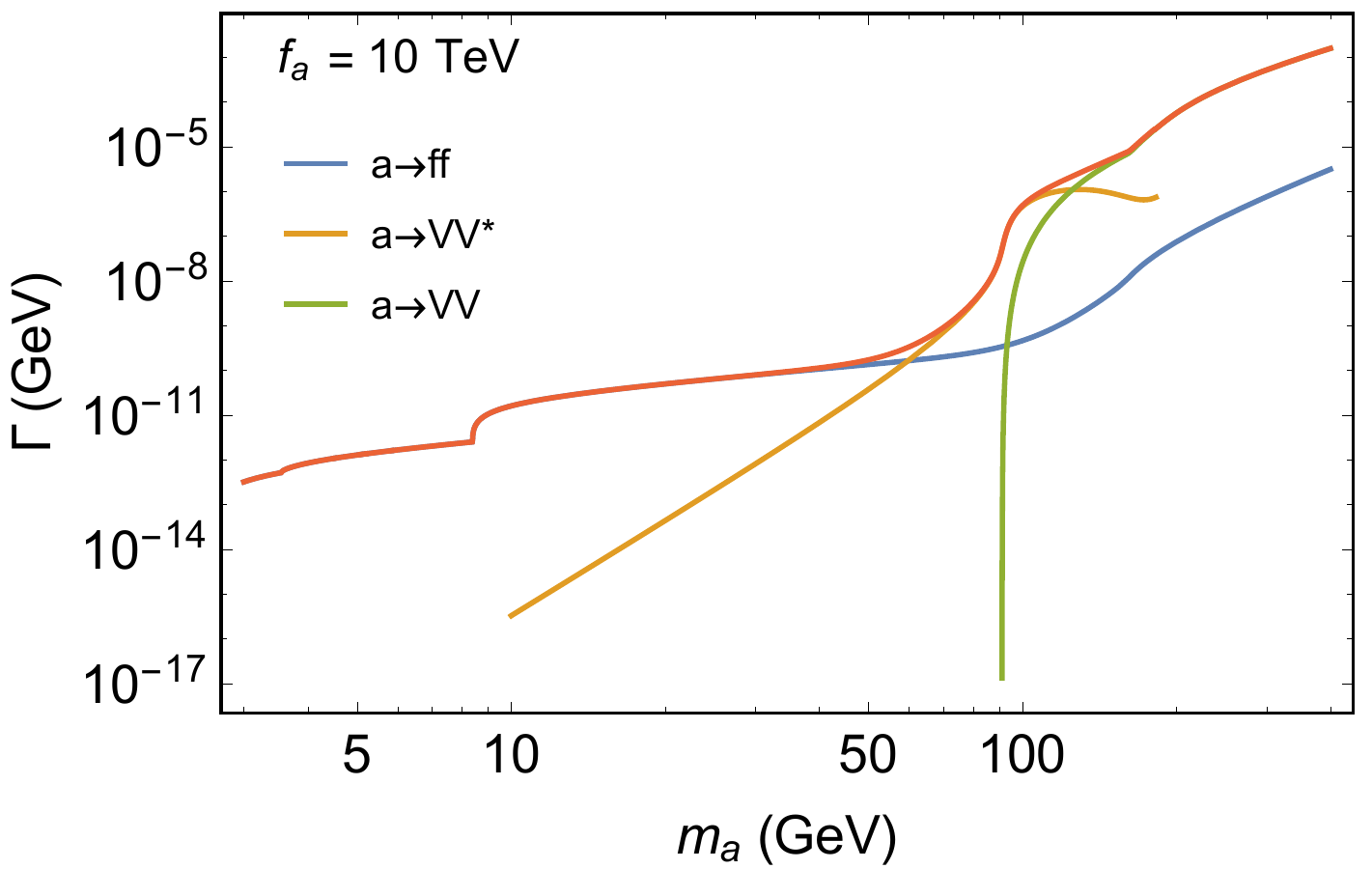} 
   \caption{Partial decay widths of the photophobic ALP as a function of mass. Tree level decays to pairs of electroweak gauge bosons dominate whenever kinematically accessible, and three-body decays via off-shell gauge bosons remain significant below threshold for pair production.  Loop induced decays to fermions dominate when the axion is significantly lighter than the massive electroweak gauge bosons. Below $2m_e$ (not shown), decays to photon pairs finally prevail.}
   \label{fig:gamma}
\end{figure}

\subsection{ALP Production}

The tree-level couplings of the photophobic ALP to pairs of electroweak gauge bosons also dominate its production at lepton and hadron colliders. The most relevant production modes include $\gamma a$, $Z a$, and $W^\pm a$ associated production.  

\begin{figure}[htbp] 
   \centering
   \includegraphics[width=5in]{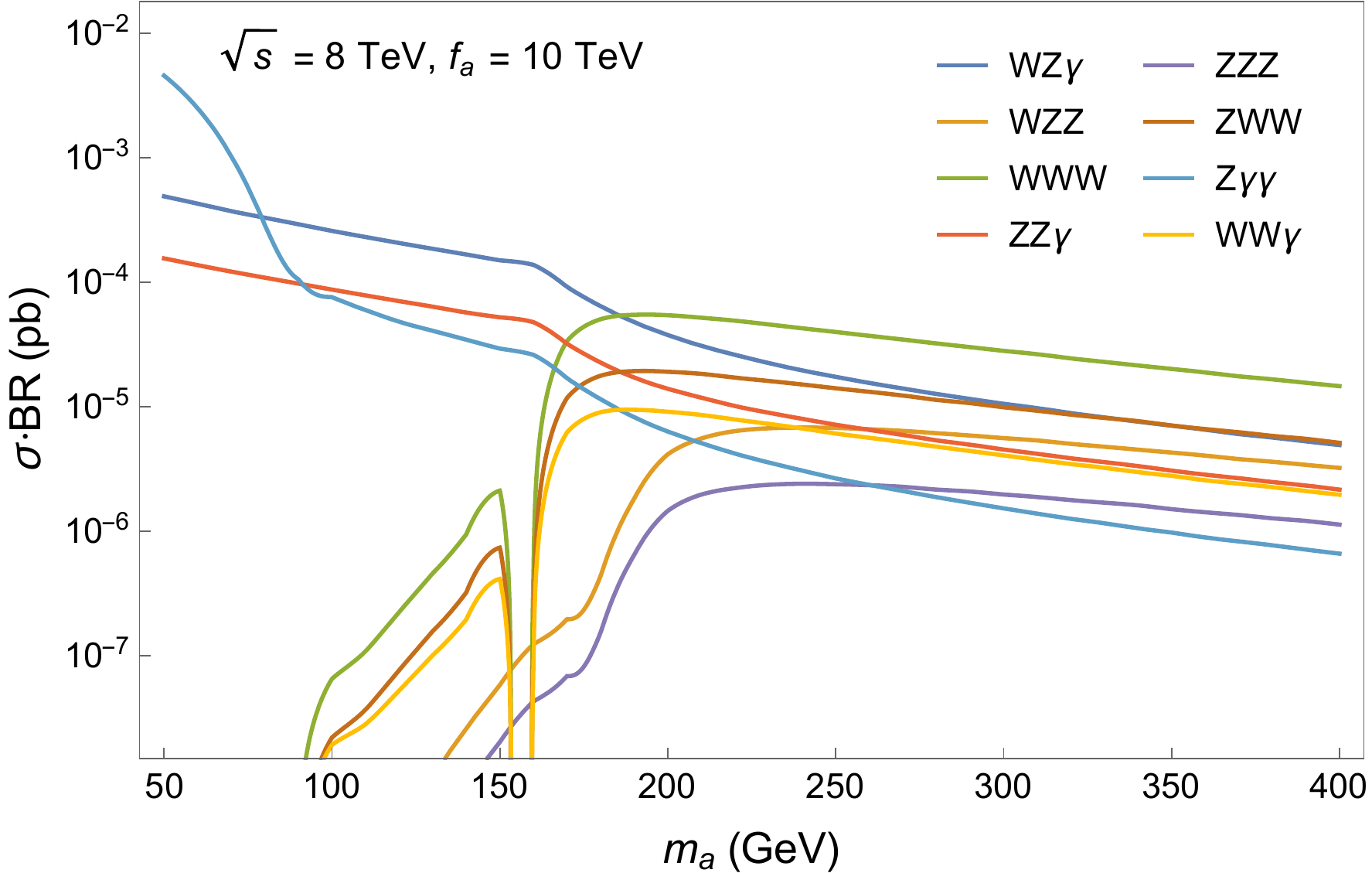} 
   \caption{Leading order cross sections times branching ratios for multi-boson final states produced via a photophobic ALP at the LHC. Note that these final states combine different associated production and decay modes of the photophobic ALP, such as $Z a \rightarrow ZZ\gamma$ and $\gamma a \rightarrow ZZ \gamma$. Rates for $m_a \lesssim m_Z$ include contributions from off-shell massive vectors. Here we have taken $\sqrt{s} = 8$ TeV and $f_a = 10$ TeV. }
   \label{fig:sigmabr}
\end{figure}

In conjunction with the dominant decays, these associated production modes lead to a wide array of collider signals that distinguish the photophobic ALP from its generic counterparts. At LEP, the dominant production mode for $m_a < m_Z$ is $\gamma a$ via an on-shell $Z$ boson, making searches for exotic decays of the form $Z \rightarrow \gamma + X$ particularly sensitive probes of the photophobic ALP. At hadron colliders such as the LHC, associated production of higher-mass photophobic ALPs leads to a rich set of triboson final states such as $WZ \gamma$, $WZZ$, $WWW$, $ZZ\gamma$, $ZZZ$, $ZWW$, $Z \gamma \gamma$, and $WW \gamma$. Consequently, many of the best existing LHC constraints on photophobic ALPs come from Standard Model triboson measurements. The rates for these triboson processes are illustrated in Figure~\ref{fig:sigmabr} at the $\sqrt{s} = 8$ TeV LHC, for which the most comprehensive set of triboson measurements is available. Constraints on the photophobic ALP from existing LEP and LHC searches are explored in the next section.

\section{Photophobic Limits}
\label{sec:limits}

The parameter space of the photophobic ALP is constrained by a variety of astrophysical and terrestrial systems.\footnote{We do not include potential cosmological constraints, as in the region of parameter space not already constrained by astrophysical and terrestrial probes, it is possible for the reheating temperature to be sufficiently low so as to avoid producing a significant cosmological population of photophobic ALPs while remaining consistent with BBN.} The most significant astrophysical constraints come from anomalous cooling of red giants via the ALP coupling to electrons and anomalous energy loss in Supernova 1987A via the ALP coupling to nucleons. Terrestrial constraints arise primarily from rare meson decays probed in a variety of experiments and, at higher mass, direct searches at LEP and the LHC. The most relevant constraints on photophobic ALPs in the eV -- TeV range are summarized in Figure \ref{fig:AllLims} and discussed in detail below.

\begin{figure}[h]
	\centering
	\includegraphics[width=5in]{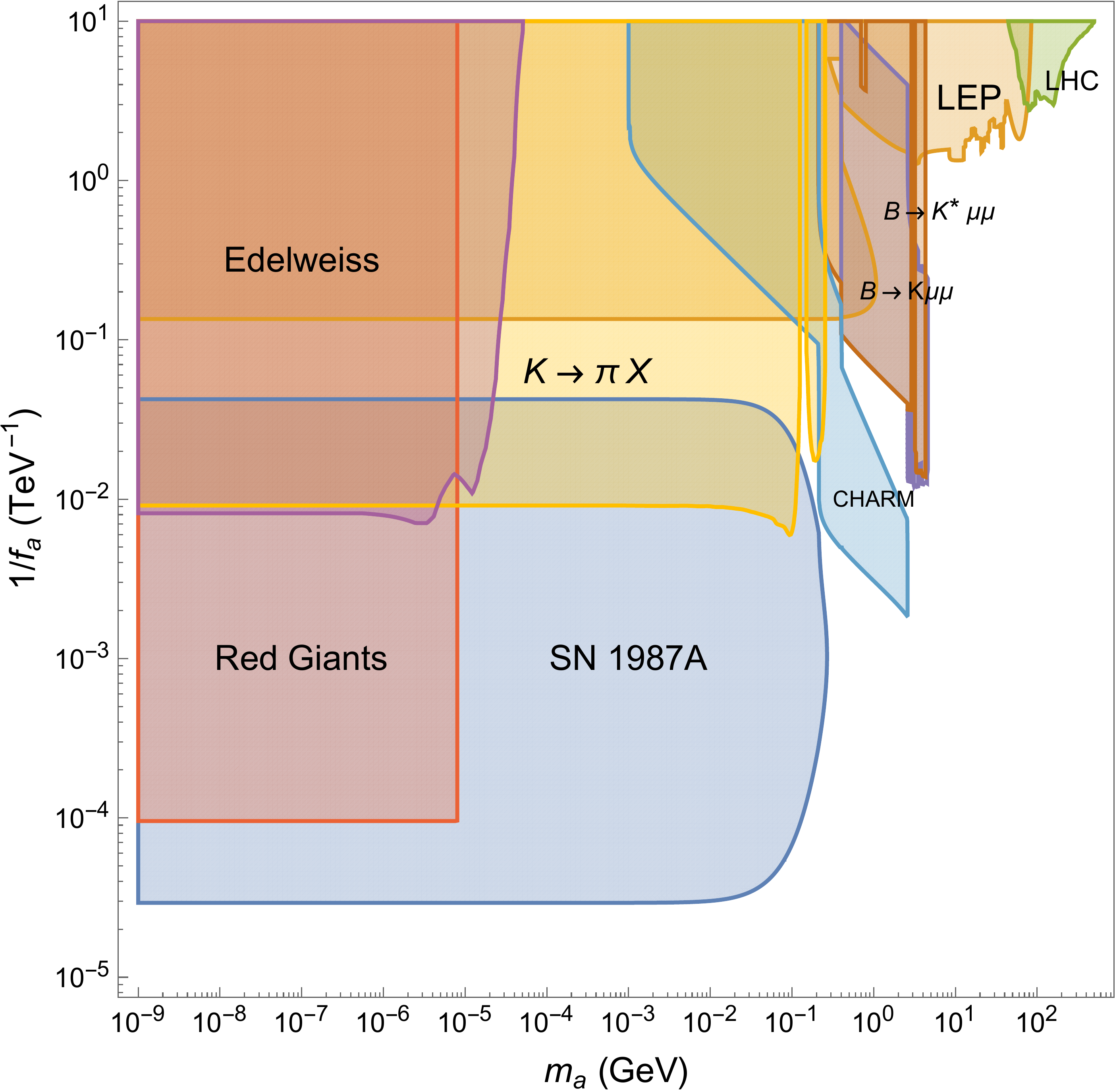}
	\caption{Summary of relevant limits on the mass $m_a$ and coupling $f_a$ of the photophobic ALP from astrophysical and terrestrial probes. Here we have taken $\Lambda = 10$ TeV.}
	\label{fig:AllLims} 
\end{figure}

\subsection{Red Giants}

The coupling of ALPs to electrons can be constrained by observations of red giant stars, in which cooling due to processes such as bremsstrahlung $e + Ze \rightarrow Ze + e + a$ leads to larger core masses at the time of helium ignition. Relatively good agreement between observation and theoretical expectations for the core mass at helium ignition constraints the effective coupling between ALPs and electrons for ALP masses below the typical red giant core temperature $T \approx 10^8$ K, i.e. $m_a \lesssim 9 \times 10^{-6}$ GeV \cite{Raffelt:1994ry}. The loop-induced coupling of the photophobic ALP to electrons is roughly two orders of magnitude smaller than the coupling to gauge bosons, constraining $f_a \gtrsim 10^4$ TeV in the mass range of interest. 

\subsection{Edelweiss}

Null results in searches for ALPs at the EDELWEISS-II experiment \cite{Armengaud:2013rta} lead to constraints on the coupling of ALPs to photons, the coupling of ALPs to electrons, and the product of couplings to electrons and nucleons. In each case, the constraints rely only on axion production in the sun, rather than assuming a cosmological abundance of dark matter axions.  The strongest constraints on the photophobic ALP come from solar Compton, bremsstrahlung, axio-recombination, and axio-deexcitation processes, bounding the product of ALP couplings to electrons and nucleons. For photophobic ALPs below 10 keV this corresponds to a bound $f_a \gtrsim 120$ TeV. Constraints on the ALP coupling to electrons coming from 14.4 keV axions emitted in the M1 transition of $^{57}$Fe nuclei are roughly an order of magnitude weaker than the above constraints for the photophobic ALP. Constraints on the ALP coupling to photons are irrelevant for the photophobic ALP due to the small induced photon coupling in the mass range of interest. 

\subsection{Supernova 1987A}

The ALP coupling to photons or nucleons is famously bounded by the neutrino signal observed from SN 1987A, which constrains anomalous cooling due to the Primakoff process $\gamma + p \rightarrow p + a$ or the nucleon bremsstrahlung process $N + N \rightarrow N + N + a$ for ALP masses beneath about 100 MeV. For the photophobic ALP the dominant constraint is due to nucleon bremsstrahlung. We compute the anomalous energy loss rate due to nucleon bremsstrahlung following \cite{Raffelt:1996wa}, requiring the total energy outflow for SN 1987A to not exceed $\sim 3 \times 10^{53}$ erg. 

Of course, for sufficiently large couplings the ALPs become trapped in the supernova core and do not lead to efficient energy loss. Since ``efficient'' is only determined relative to energy loss due to neutrinos, the relevant comparison is not between the ALP mean free path and the supernova core size, but rather between the ALP opacity and the neutrino opacity. We compute the photophobic ALP Rosseland mean opacity $\kappa_a$ due to ALP-nucleon interactions following \cite{Raffelt:1996wa} and impose the SN 1987A energy-loss bound only when $\kappa_a < \kappa_\nu$, where $\kappa_\nu \approx 8 \times 10^{-17}$ cm$^2$/g is the neutrino opacity. This leads to the bound shown in Figure~\ref{fig:AllLims}. We have verified that energy losses due to $\gamma$- or $Z$-mediated Primakoff processes are relevant only in regions where the photophobic ALP mean opacity significantly exceeds the neutrino opacity.

\subsection{Rare meson decays}
Through its coupling to $W^\pm$ bosons, the photophobic axion mediates rare flavor-changing neutral current (FCNC) processes. The strongest limits can be obtained from, e.g.,  bounds on $K^\pm \to \pi^\pm + X, B^\pm \to K^\pm + X$, and $B^0 \to K^{0*} + X$.

Bounds on the branching ratio $K^+ \to \pi^+ a$ for invisible $a$ have been calculated by the E949~\cite{Artamonov:2009sz} collaboration, and interpreted as a bound on low-mass ALPs coupled to $W$ bosons in~\cite{Izaguirre:2016dfi}.  In the future, NA62 will place a more constraining limit on the branching ratio, though due to an upward fluctuation of their data they do not currently place the tightest constraints~\cite{NA62}. In the mass range of interest, the photophobic ALP is typically sufficiently long-lived to be effectively invisible. We have correspondingly reinterpreted the limit using the original E949 analysis and find the bounds shown in Figure~\ref{fig:AllLims}.

For $m_a > 2 m_\mu$, the decay length of $a$ falls beneath 100 cm for $f_a \lesssim 10$ TeV, so that the strongest limits come from rare $B$ decays in which the ALP subsequently decays to a pair of muons. The LHCb collaboration sets exceptionally strong bounds on both $B^\pm \rightarrow K^\pm + a$ \cite{Aaij:2016qsm} and $B^0 \rightarrow K^{0*} + a$ \cite{Aaij:2015tna} with $a \rightarrow \mu^+ \mu^-$ as a function of the $a$ mass and lifetime. We directly reinterpret the two-dimensional limits presented in \cite{Aaij:2016qsm,Aaij:2015tna} to obtain the corresponding bounds shown in Figure~\ref{fig:AllLims}. 

Proton beam dump experiments looking for long-lived particles set a constraint complementary to those coming from direct searches for rare meson decays. Production occurs via the rare decays $K^\pm \rightarrow \pi^\pm + a, K_L \rightarrow \pi^0 + a$, and $B \rightarrow K + a$, followed by the displaced decay $a \rightarrow \gamma \gamma, e^+ e^-, \mu^+ \mu^-$ in the detector. The strongest bound comes from the CHARM experiment \cite{Bergsma:1985qz}, for which we use the reinterpretation framework of \cite{Clarke:2013aya} to set the bound shown in Figure~\ref{fig:AllLims}. Limits set by electron beam dump experiments are far weaker. Rare $\Upsilon$ decays proceed via an off-shell $Z$ and likewise lead to subdominant limits.

The unusual shape of the bound coming from the CHARM experiment is due to a pair of effects. The first is that the lifetime of the ALP changes sharply at twice the muon mass as decays into muon pairs open up. This leads to the kink in the limit for 200 MeV $< m_a <$ 350 MeV, where the limit on $f_a$ changes rapidly as a function of $m_a$. The far boundary of the kink around 350 MeV arises because the dominant limit is coming from kaon decays into an ALP and a pion, which shuts off above 350 MeV. The second feature is due to the modest contribution from B meson decays at higher masses, which give an additional contribution at higher $m_a$ and small $f_a$.

\subsection{LEP}

LEP provides the leading constraint on the photophobic ALP over a range of masses thanks to a combination of complementary searches. ALPs between $\sim 1-100$ GeV decay promptly and are best constrained by a combination of direct searches and the model-independent constraint on the $Z$ width, while the displaced decays of lighter ALPs are constrained by searches for partly invisible decays of the $Z$. The leading bounds are summarized in Figure~\ref{fig:LEPLims}.

\begin{figure}[t]
	\centering
	\includegraphics[width=8cm]{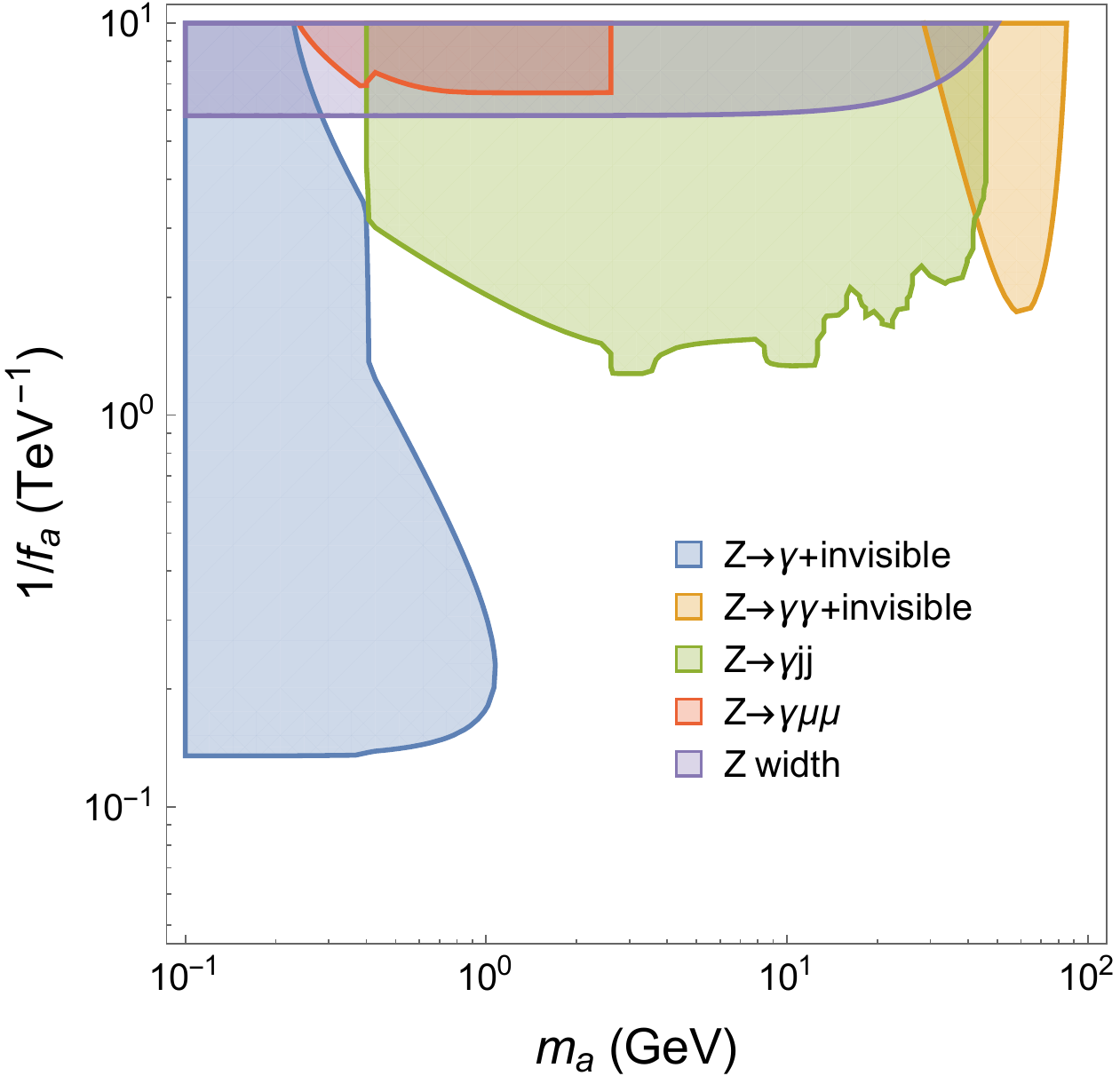}
	\caption{LEP limits on the photophobic ALP from searches for $Z \rightarrow \gamma$ + invisible, $Z \rightarrow \gamma \gamma$ + invisible, $Z \rightarrow \gamma (jj)$, $Z \rightarrow \gamma \mu^+ \mu^-$, and the model-independent bound on new physics contributions to the $Z$ width. Here we have taken $\Lambda = 10$ TeV.}
	\label{fig:LEPLims} 
\end{figure}

When the photophobic ALP decays promptly, a significant limit arises at higher masses from searches for $Z\to \gamma \gamma + {\rm invisible}$, most notably from from an OPAL search at the $Z$ pole~\cite{Acton:1993kp} that constrains
\begin{equation}\label{eq:BRjjgg}
BR(Z\to \gamma \gamma + {\rm invisible}) < 3.1 \times 10^{-6}
\end{equation}
Photophobic ALP contributions to $Z\to \gamma \gamma + {\rm invisible}$ from a promptly-decaying ALP arise primarily from events with $Z \rightarrow \gamma a \rightarrow \gamma \gamma \nu \bar \nu$. Properly speaking, the above OPAL analysis focuses on a diphoton invariant mass window of $55 < m_{\gamma \gamma} < 65$ GeV, but sees zero signal events for $m_{\gamma \gamma} \gtrsim 10$ GeV. As such, we set a limit assuming the acceptance for events with $m_{\gamma \gamma} \gtrsim 10$ GeV is comparable to the acceptance quoted for $55 < m_{\gamma \gamma} < 65$ GeV. This provides the leading limit for $50 \, {\rm GeV} \lesssim m_a \lesssim m_Z$.

At lower masses, promptly-decaying photophobic ALPs are predominantly constrained by an L3 search on the $Z$ pole for $Z \rightarrow \gamma a \rightarrow \gamma jj$ \cite{Adriani:1992zm}. This provides the leading LEP constraint for $12  \lesssim m_a \lesssim 50$ GeV. A somewhat weaker bound arises from the model-independent LEP limit on anomalous contributions to the $Z$ width \cite{ALEPH:2005ab}, which constrains $\Gamma_{\rm new} < 2.3$ MeV at $1 \sigma$ and provides coverage at lower masses beneath the acceptance of the $Z \rightarrow \gamma a \rightarrow \gamma jj$ search. A potential bound also arises from limits on $Z \rightarrow \gamma \mu^+ \mu^-$ \cite{Acton:1991dq}, but this is ultimately a subleading limit in the region of interest.
 
At still lower masses, the ALP becomes long-lived and eventually susceptible to LEP searches for $Z \rightarrow \gamma + {\rm invisible}$. An L3 search in $e^+ e^-$ collisions at the $Z$ resonance at LEP constrains the branching ratio $Z\to \gamma X$ for stable $X$ by~\cite{ACCIARRI1997201}
\begin{equation}
BR(Z\to \gamma X) < 1.1\times 10^{-6}
\end{equation}
where the energy of the photon is greater than 31 GeV. To set a limit in this case, we simulate $e^+ e^- \rightarrow Z \rightarrow \gamma + a$ events in {\texttt MadGraph} for a range of $m_a$, impose the requirement that $E_\gamma > 31$ GeV, and count as invisible all events for which the ALP travels a distance of at least 100cm before decaying. This provides the strongest LEP constraint at lower masses, though it is ultimately superseded by bounds from rare meson decays.

\subsection{LHC}

The LHC is sensitive to higher-mass photophobic ALPs, particularly through the rich set of triboson final states accessible for $m_a > m_Z$. In order to obtain LHC limits on the photophobic ALP, we consider LHC searches probing any of the triboson final states in Figure~\ref{fig:sigmabr} with potential sensitivity to a photophobic ALP signal. For the masses and coupling strengths of interest, the most relevant searches are Standard Model measurements of triboson production. Of these, searches for the $Z\gamma \gamma$ final state give the best limits of any of the LHC analyses considered over the entire ALP mass range. Although the $WWW$ and $WZ\gamma$ final states have the highest values of cross section times branching ratio for higher ALP masses, followed by other $W$- and $Z$-associated production modes, the higher signal efficiencies of the LHC $Z\gamma \gamma$ analyses lead to stronger limits.

\begin{figure}[t]
	\centering
	\includegraphics[width=8cm]{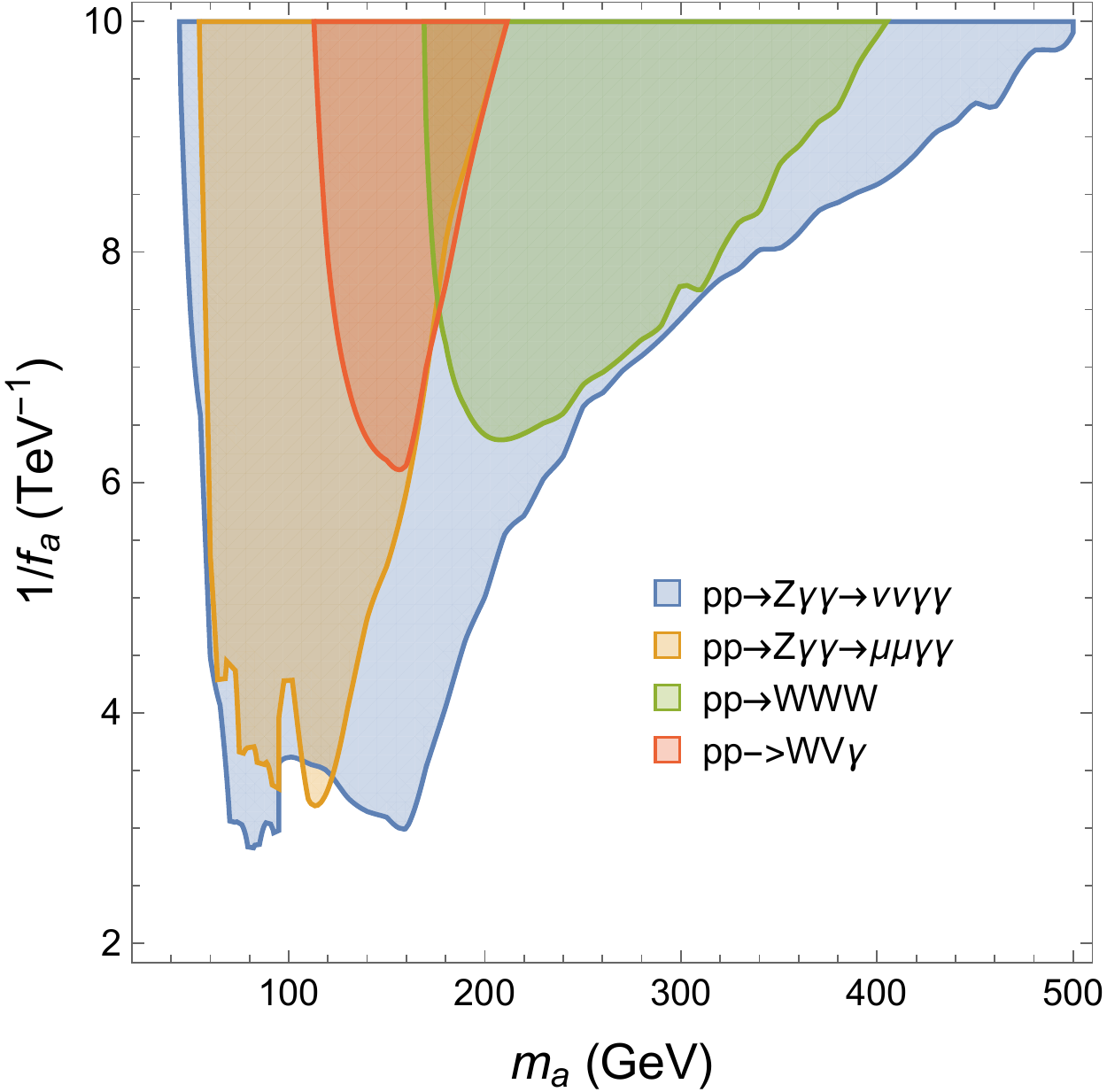}
	\caption{LHC bounds on the photophobic ALP at 95\% CL from reinterpretations of triboson searches for $pp \rightarrow Z \gamma \gamma \rightarrow \nu \bar \nu \gamma \gamma$, $pp \rightarrow Z \gamma \gamma \rightarrow \mu^+ \mu^- \gamma \gamma$, $pp \rightarrow WWW$, and $pp \rightarrow WV \gamma$ at $\sqrt{s} = 8$ TeV. }
	\label{fig:LHCLims} 
\end{figure}

The most sensitive $Z\gamma\gamma$ search is the ATLAS Standard Model $Z\gamma$ and $Z\gamma\gamma$ search with leptonic decays of the $Z$ at $\sqrt{s} = 8$ TeV~\cite{Aad:2016sau}. For each of the three final states of the $Z$ included in the search, $\nu\bar \nu$, $\mu^+ \mu^-$, and $e^+e^-$, the best-discriminating bin of the $m_{\gamma \gamma}$ distribution was used to set a Poisson limit on an anomalous $Z \gamma \gamma$ signal. Then a simulation sample of the ALP-mediated process was generated in MadGraph and used to bound the ALP coupling constant $1/f_a$ (see Appendix~\ref{sec:limitappZGG} for details). The electron decay channel was less sensitive than the muon channel throughout the mass range. Results for both the $\nu\bar \nu$ and $\mu^+ \mu^-$ final states are shown in Figure~\ref{fig:LHCLims}.

 For the $WWW$ final state, the ATLAS Standard Model $W^\pm W^\pm W^\mp$ search at $\sqrt{s} = 8$ TeV~\cite{Aaboud:2016ftt} gives the best sensitivity. This search includes six analysis channels for various decays of the $W$s: $\ell^\pm \nu \ell^\pm \nu \ell^\mp \nu$ and  $\ell^\pm \nu \ell^\pm \nu jj$ with $\ell = e$ or $\mu$. First, limits from all six analysis channels were approximated  by taking the signal efficiencies to be the same as Standard Model $WWW$ efficiencies, and the relative signal rates to be proportional to the Standard Model signal rates in each channel. Once it was determined that the best limit should come from the $\mu^\pm \nu \mu^\pm \nu j j$ signal region, the event yield and predicted background for that region were used to set a Poisson limit on an anomalous $WWW$ signal. Once again a simulation sample of the ALP-mediated process was used to bound the ALP coupling constant (details in Appendix~\ref{sec:limitappWWW}). Bounds from the CMS $WH$ with $H \to W^+ W^-$ search~\cite{CMS:2013xda} were also calculated, but are worse than those from the ATLAS Standard Model search throughout the mass range.
 
 Finally, the search with the most sensitivity in the $WV\gamma$ final states is the CMS Standard Model $WW\gamma$ and $WZ\gamma$ search at $\sqrt{s}=8$ TeV~\cite{Chatrchyan:2014bza}. 
 The published limit on cross section times branching ratio of 311 fb was compared with a simulation sample to set a limit on the ALP coupling (details in Appendix~\ref{sec:limitappWVG}).
 
Various BSM searches were considered as well, but simple rate estimates suggest that none of these analyses provide comparable sensitivity to a photophobic ALP signal. Limit curves for the most sensitive Standard Model searches are shown in Figure~\ref{fig:LHCLims}. Although our reinterpretation of existing Standard Model triboson measurements set significant limits on photophobic ALPs between 40--500 GeV, it is highly likely that dedicated triboson searches could significantly extend limits in this range. In this respect, photophobic ALPs motivate the expansion of low-mass multi-boson searches at the LHC.

\section{Photophobic Relaxion}
\label{sec:relaxion}

\newcommand\eea{\end{eqnarray}}
\newcommand\bea{\begin{eqnarray}}

We now briefly discuss how our results affect relaxion models which utilize photophobic axions.
Particle production relaxation takes a step towards solving some of the problems of the relaxion approach to the electroweak hierarchy problem.
Many of these problems such as super planckian field excursions, low scale inflation, and large amounts of inflation, are tied to fact that
relaxation occurs during inflation.  Particle production relaxation divorces the relaxion from inflation by using particle production as a frictional force instead of Hubble expansion.  Instead, relaxation occurs after inflation, removing any constraints on inflation.  Additionally, the field excursions can be sub-planckian.

In the relaxion approach, a small weak scale is selected due to the appearance of minima around the weak scale.  In particle production relaxation,
there are always minima, but a frictional force appears at the small weak scale.  Only when the electroweak gauge bosons are light enough that they can be 
produced do they provide a friction term for the relaxion, thereby selecting a vacuum where the Higgs boson is light.  In this approach, the relaxion 
necessarily dominantly couples to the electroweak gauge bosons.  Any coupling to the photon, provides a Higgs independent friction term that could 
potentially ruin the mechanism.  More details can be found in Ref.~\cite{Hook:2016mqo}.

This work adds two major constraints to these types of models.  The first constraint is that 
there is necessarily a coupling of the relaxion to the photon and it needs to be checked that this new source of friction is small.
Requiring that the timescale for photon production be longer than Hubble gives the constraint
 \bea
 \label{Eq: fphoton}
 f_{\gamma \gamma} \gtrsim \frac{v f_a}{H}
 \eea
However, as emphasized earlier in the paper, a coupling to photons cannot be completely suppressed.  At two loops, via a loop of electrons, the relaxion obtains an unavoidable
coupling to the photon that is parametrically $f_\gamma \sim f_a m_e^2/m_a^2$.  Requiring that this two loop coupling satisfy the bound shown in Eq.~\ref{Eq: fphoton} is a non-trivial constraint.
The second constraint is that the one loop coupling to fermions is bounded by astrophysical constraints discussed in Section \ref{sec:limits}.  These constraints are non-trivial.  For example, the original data point shown in Ref.~\cite{Hook:2016mqo} has a two loop coupling to the photon that is too large for the mechanism to work.

These new constraints place particle production relaxation in an interesting position.  If the relaxion is too weakly coupled, then particle production will be too small of an effect to slow the relaxion down.  If the relaxion is too strongly coupled, it runs afoul of the astrophysical constraints.  The result of the tension between these two effects is that typical data points which satisfy all of the criteria to make particle production relaxation work are in mild tension with astrophysical constraints.
The following data point is an an example of a typical set of parameters which satisfy this constraint: The relaxion has a mass of keV, $f_a = 10^6$ GeV and an induced coupling to
photons of $10^{16}$ GeV.~\footnote{For those more familiar with these models, the parameters of the theory are as follows:
The UV cutoff is $\Lambda = 10^4$ GeV, 
the linear term for the relaxion is $\epsilon = 10^{-8}$ GeV, $f_a = 10^6$ GeV, the confinement scale is $\Lambda_c = 10^3$ GeV and the coupling to the confining sector is $f' = 10^{12}$ GeV.}  Note that an $f_a$ this small is in tension with the astrophysical constraints discussed in Section \ref{sec:limits} where $f_a \gtrsim $ few $10^7$ GeV.  The error bars typically associated with constraints coming from astrophysical sources is about an order of magnitude, leading to mild tension between the astrophysical bounds and the particle production relaxion models.

As mentioned before, the relaxion has couplings to the electroweak gauge bosons that are only suppressed by $f \sim 10^3$ TeV.  In the high temperature early universe, the relaxion condensate is destroyed
and a thermal population of relaxions is produced.  If present, the thermal population of relaxions would overclose the universe unless they decayed.  As a result, we introduce a coupling of the relaxion to lighter particles, e.g. right handed neutrinos, that allows for the relaxion to dump its energy into lighter particles.
The energy density in the relaxion or its decay products can be tested as it is an additional contribution to $N_\text{eff}$ at BBN that is $\mathcal{O}(0.1)$.

In summary, the one loop induced coupling to fermions and the two loop induced coupling to photons place particle production relaxation in a very precarious situation.  The viable regions of parameter space of the relaxion are in mild conflict with astrophysical constraints and require that the actual bounds be about an order of magnitude weaker than those estimated in this paper.  These concerns warrant a more careful investigation of these models.

\section{Conclusions}
\label{sec:conclusion}

In this paper, we have demonstrated the extent to which an ALP can be photophobic.  While photophobia may appear to be a fine tuning, we give an explicit UV construction where the lack of coupling to photons is in fact due to a left-right symmetry.  This UV construction realizes maximal photophobia for the ALP.
Using symmetry-based arguments, we then argued that a photophobic axion necessarily has an irreducible coupling to the photon that is proportional to the mass of the ALP, a fact supported by explicit calculation of radiative corrections. The dominant coupling to the photons for low mass photophobic ALPs is a two loop effect mediated by the electron.

The suppression of the photophobic ALP coupling to photons is sufficient to drastically change the constraints on the ALP relative to its generic counterparts.  At low masses, the dominant constraints are astrophysical in nature and weaker than the typical astrophysical bounds due to the smallness of radiative photon/fermion couplings.  For ALPs in the MeV to GeV range, the dominant constraints are due to rare meson decays.  Photophobic ALPs with a mass above a GeV are the most distinct from their photophilic ALP counterparts, as the suppressed diphoton decay mode opens up a range of alternative final states.  Precision LEP measurements of the Z boson constrain ALPs with masses between a GeV and 100 GeV.  Above 100 GeV, LHC triboson searches place the dominant constraint. While limits currently arise from reinterpretations of Standard Model triboson measurements, there is significant room for improvement via dedicated triboson searches for photophobic ALPs.

The presence of an irreducible coupling to the photon even for photophobic axions has implications for particle production relaxion models aimed at addressing the electroweak hierarchy problem.  In particular, the irreducible photon coupling of photophobic ALPs places the simplest models of particle production in mild tension with constraints.  It would be interesting to further explore such models in detail to see how robustly constrained they truly are. \\ 

\noindent {\bf Note added:} After this work was completed, \cite{Fonseca:2018xzp} appeared exploring post-inflationary particle production relaxion models involving photophobic ALPs. 

\acknowledgments

We thank Diego Redigolo, Gilad Perez, and Sophie Renner for helpful conversations. NC and SK are supported in part by the US Department of Energy under the grant DE-SC0014129.  AH is supported  Grant No. NSF PHY-1620074 and by the Maryland Center for Fundamental Physics (MCFP). We thank the Kavli Institute of Theoretical Physics for hospitality during the completion of this work, and corresponding support from the National Science Foundation under Grant No. NSF PHY-1748958.

\appendix
\section{Photophobic Models}
\label{app:models}

\newcommand{\be}{\begin{equation}}
\newcommand{\ee}{\end{equation}}
\newcommand\eea{\end{eqnarray}}
\newcommand\bea{\begin{eqnarray}}

In this Appendix, we give a simple left right (LR) symmetric model that gives a photophobic axion.  The matter content of the theory is
\begin{center}
\be
\begin{tabular}{c|c|cccc|c}
&$SU(N_c)$&$SU(2)_L$&$SU(2)_R$&$SU(3)$&$U(1)$&$U(1)_a$\\
\hline
&&\\[-8pt]
$Q_L$ &  & $2$ & & $3$ & $1/6$ &  \\
$Q_R$ &  &  & $\overline 2$ & $ \overline 3$ & $-1/6$ &  \\
$L_L$ &  & $2$ & & & $-1/2$ &  \\
$L_R$ &  & & $\overline 2$ & & $1/2$ &  \\
$H_L$ &  & $2$ & & & $1/2$ &  \\
$H_R$ &  & & $\overline 2$ & & $-1/2$ &  \\
\hline
$q_L$ & $N_c$ & $2$ & &&& 1 \\
$q_L^c$ & $\overline N_c$ & $\overline 2$ & &&& 1 \\
$q_R$ & $N_c$ & & $2$ &&& -1\\
$q_R^c$ & $\overline N_c$ & & $\overline 2$ &&& -1
\end{tabular}
\ee
\end{center}
where the first set of particles are the SM particles with flavor indices suppressed and the second set are the confining quarks whose pseudogoldstone boson will be our axion.  There is an additional $SU(N_c)$ confining gauge group.  The accidental anomalous global symmetry which will give the axion is denoted $U(1)_a$.

Ignoring the Standard Model part of the Lagrangian, the leading order renormalizable Lagrangian is
\bea
V = \epsilon (q_L q_L^c + q_R q_R^c)
\eea
The small explicit mass terms for $q$ breaks the anomalous symmetry and gives the axion a mass.  Under the left right symmetry of the theory
\bea
Q_L \leftrightarrow Q_R^\dagger \qquad q_L \leftrightarrow q_R^{c,\dagger}
\eea

When the $SU(N_c)$ confines with chiral symmetry breaking, there is an IR theory of pseudo-Goldstone bosons.  In the chiral Lagrangian, the axion is
\bea
 \begin{bmatrix}
    q_L q_L^c       & q_L q_R^c  \\
    q_R q_L^c       & q_R q_R^c 
\end{bmatrix}
 =  \begin{bmatrix}
    1       & 0  \\
    0      & 1
\end{bmatrix} e^{i \frac{a}{f_a}  \begin{bmatrix}
    1       & 0  \\
    0      & -1
\end{bmatrix}}
\eea
The axion obtains a mass
\bea
m_a^2 \sim \epsilon \Lambda_{N_c} \qquad f_a \sim \frac{\Lambda_{N_c}}{4 \pi} 
\eea
The axion is CP odd and under the LR symmetry is even.  These two symmetry properties will result in the axion having the properties needed to be photophobic.

Using direct calculations or appealing to the symmetry properties of the axion, one finds that the axion has no fermion couplings and only couples to
\bea
\label{Eq: LR coupling}
\frac{g^2 a}{32 \pi^2 f_a} (W_R \tilde W_R - W_L \tilde W_L)
\eea
This results in the axion having the anomalous shift symmetry
\bea
a \rightarrow a + \epsilon f_a \qquad \theta_L \rightarrow \theta_L + \epsilon \qquad \theta_R \rightarrow \theta_R - \epsilon
\eea
that is only broken by its mass.  The spurions $\theta_R$ and $\theta_L$ are topological so that regardless of how left-right symmetry is broken, we know that this symmetry descends into the anomalous symmetry shown in Eq.~\ref{Eq: symmetry}.

The statement that the axion has no fermion couplings and couples only as in Eq.~\ref{Eq: LR coupling} is not a coincidence and is enforced by the symmetries of the theory.  The additional couplings to gauge bosons
$G \tilde G$ , $B \tilde B$ and  $W_L \tilde W_L + W_R \tilde W_R$ are all odd under LR symmetry and thus cannot couple to the axion.  The fermion couplings are either both CP and LR even or both CP and LR odd.  Thus
the axion cannot couple to the fermions.  A simple trick to see the lack of coupling to fermions is to use a field redefinition to change the current coupling of the axion into a mass term coupling of the axion $m_f e^{i a} f f^c$.
Parity sets
\bea
Y_u = Y_u^\dagger \qquad Y_d = Y_d^\dagger
\eea
Because the axion carries no flavor indices and can only couple to the overall $Y_u$, which is hermitian, we see that the axion cannot couple to the standard pseudoscalar fermion currents.

LR and CP symmetry properties of the axion prevent it from coupling to any of the other gauge bosons or fermions.  After LR symmetry is broken, RG evolution will reintroduce the coupling to the fermions.  Thus the UV scale in Eq.~\ref{Eq: fermions} is the scale of the LR symmetry breaking.  As LR symmetry breaking is independent of the shift symmetry of the axion, the shift symmetry still prevents couplings of the axion to the photon up to corrections due to the shift symmetry breaking mass of the axion.  Thus this theory provides a UV completion where the axion does not couple to the fermions or photons in the UV and all other couplings are generated in the IR due to RG evolution as described in Sec.~\ref{sec:spurion}.

\section{LHC Searches for Photophobia}
\label{app:limitapp}

Here we summarize our reinterpretation of various Standard Model multi-boson searches at ATLAS and CMS to set bounds on the photophobic ALP.
	
	Monte Carlo simulation samples were generated with the MadGraph 5.2.5.2 event generator~\cite{Alwall:2014hca} using the CTEQ6L1 parton distribution functions~\cite{Pumplin:2002vw}. 
	MadGraph was interfaced with the PYTHIA 6.428 program~\cite{Sjostrand:2006za} 
	for hadronization and underlying event simulation and DELPHES 3.3.3~\cite{deFavereau:2013fsa} for detector simulation. The resulting output was analyzed with MadAnalysis 5.1.5~\cite{Conte:2012fm,Conte:2014zja}.
	\subsection{ATLAS $Z\gamma$ and $Z\gamma\gamma$}
	\label{sec:limitappZGG}
	This search in $pp$ collisions at $\sqrt{s} = 8$ TeV used a data sample with an integrated luminosity of $\mathcal L = 20.3$ fb$^{-1}$~\cite{Aad:2016sau}. We consider the two production channels that proved most sensitive to an ALP signal, $pp\to \nu\bar\nu \gamma\gamma$ and $pp\to \mu^+\mu^- \gamma\gamma$.
	
	For a range of values of $m_a$, matched samples 
	 of 50,000 events were generated in MADGRAPH for the process $pp\to \gamma a$, $a \to \gamma x\bar x$ with $x=\nu$ or $\mu$, and passed through Pythia 6 and Delphes. MadAnalysis was used to impose the following cuts:
	\begin{itemize}
		\item Isolation requirements:
		\begin{itemize}
			\item Photon: sum of P$_T$ in a cone $\Delta R = 0.4 < 4$ GeV
			\item Muon: sum of $P_T$ in a cone $\Delta R = 0.2 < 0.1\times ($Candidate $P_T$)
		\end{itemize}
		
		\item Cuts shared by both the $\nu\bar\nu \gamma\gamma$ and $\mu^+\mu^- \gamma\gamma$ analyses
		
		\begin{itemize}
			\item Contains at least two isolated photons
			\item Both photons satisfy $|\eta| < 2.37$, excluding the transition region $1.37 < |\eta| < 1.52$
			\item Photon-photon separation $\Delta R(\gamma,\gamma) > 0.4$
		\end{itemize}
		
		\item $\mu^+\mu^-\gamma\gamma$ cuts
		\begin{itemize}
			\item Contains exactly one pair of oppositely-charged, isolated muons
			\item Muon $|\eta| < 2.5$
			\item Muon $p_T > 25$ GeV
			\item Both photons $E_T > 15$ GeV
			\item Neither photon within $\Delta R = 0.4$ of a muon
			\item Invariant mass of muon pair $> 40$ GeV
		\end{itemize}
		
		\item $\nu\bar \nu\gamma\gamma$ cuts
		\begin{itemize}
			\item Missing E$_T > 110$ GeV
			\item Both photons E$_T > 22$ GeV
			\item Directions of diphoton system and missing $P_T$ vector separated by $\Delta \phi(\vec p_T^{\textnormal{miss}},\gamma\gamma) > 5\pi /6$
			\item No identified muons or electrons
		\end{itemize}
		
	\end{itemize}

Next, diphoton invariant mass distributions were produced with the same binnage as in the ATLAS search. For each bin, the number of events remaining after implementing all cuts was divided by the number of events in the matched sample to give the acceptance times efficiency for that bin.

Reading off the number $n$ of events observed and the expected background $b$ in each bin from the ATLAS $m_{\gamma\gamma}$ distribution in the inclusive channel (no jet requirements), 95\% confidence limits were set on the upper bound on the mean value $\mu$ of the Poisson-distribution that produced $n$ using the formula~\cite{Patrignani:2016xqp}
\[
\mu_\textnormal{up} = \frac{1}{2} F_{\chi^2}^{-1}(0.95;2(n+1)) - b
\]
where $F_{\chi^2}^{-1}$ is the quantile of the $\chi^2$ distribution. Finally, a lower limit was set on the inverse coupling strength $f_a$ using
\begin{align*}
\mu_{\textnormal{up}} &\ge \sigma \cdot \mathcal L \cdot A\times \epsilon = \sigma_{f_a = 10\textnormal{ TeV}}\left(\frac{10\textnormal{ TeV}}{f_a}\right)^2\cdot \mathcal L \cdot A\times \epsilon\\
\Rightarrow f_a &\ge (10\textnormal{ TeV})\sqrt{\frac{\sigma_{f_a = 10\textnormal{ TeV}}\cdot \mathcal L \cdot A\times \epsilon}{\mu_{\textnormal{up}}}}
\end{align*}
where the cross section $\sigma_{f_a = 10\textnormal{ TeV}}$ for a given ALP mass and $Z$ decay channel at $f_a=10$ TeV is found by multiplying the cross section for the $Z\gamma \gamma$ final state at that mass (shown in Figure~\ref{fig:sigmabr}) by the $Z$ branching fraction to that decay channel, $\mu^+\mu^-$ or $\nu\bar \nu$.

For each mass point, the limit from the single best bin in the $m_{\gamma\gamma}$ distribution is plotted in Figure~\ref{fig:LHCLims}.
	
\subsection{ATLAS $W^{\pm }W^{\pm }W^{\mp }$}
\label{sec:limitappWWW}
	
This search in $pp$ collisions at $\sqrt{s}=8$ TeV also used a data sample with an integrated luminosity of 20.3 fb$^{-1}$~\cite{Aaboud:2016ftt}. We consider the most sensitive of the six production channels, $p p \to \mu^\pm \nu \mu^\pm \nu j j$.

Simulation samples were generated as described in section~\ref{sec:limitappZGG}. The following cuts were then imposed:

\begin{itemize}
	\item Muon isolation
	\begin{itemize}
		\item sum of $P_T$ of tracks in the inner detector in a cone $\Delta R = 0.3 < 0.07 \times ($Candidate $P_T)$
		\item sum of $E_T$ in the calorimeter in a cone $\Delta R = 0.3< 0.07 \times ($Candidate $E_T)$
	\end{itemize}
\end{itemize}
	
	\begin{itemize}
		\item Exactly two same-charge muons 
		\item Both muons $p_T > 30$ GeV
		\item At least two jets satisfying $P_T^1 > 30$ GeV, $P_T^2 > 20$ GeV, and $|\eta| < 2.5$ 
		\item Invariant mass of muon pair $> 40$ GeV
		\item Invariant mass of jet pair satisfies 65 GeV $< m_{jj} < 105$ GeV
		\item Jet pair separated by $|\Delta \eta_{jj}| < 1.5$
		\item No identified electrons or additional muons with $P_T > 6$ GeV and $|\eta| < 2.5$.
		\item No identified b-jets with $P_T > 25$ GeV and $|\eta|<2.5$
		
	\end{itemize}
		
	Limits were set with a process analogous to that for the $Z\gamma \gamma$ search, using ATLAS' quoted signal + background expected value for the $\mu^\pm \nu \mu^\pm \nu j j$ signal region as our background.
	
	\subsection{CMS $WW\gamma$ and $WZ\gamma$}
	\label{sec:limitappWVG}
	This search in $pp$ collisions at $\sqrt{s}=8$ TeV used a data sample with an integrated luminosity of 19.3 fb$^{-1}$~\cite{Chatrchyan:2014bza}.
	
	Simulation samples for the $WV\gamma$ final state were generated as detailed above, and the following cuts imposed. Isolation parameters of the CMS analysis were not specified, so the default values in the Delphes CMS card were used.
	
	\begin{itemize}
		\item Contains at least one isolated photon
		\item Photon $P_T > 30$ GeV
		\item Photon $|\eta| < 1.44$
	\end{itemize}

	The CMS search found an upper limit of 311 fb at 95\% confidence level on the total cross section for $WV\gamma$ production with photons satisfying the cuts listed above. We subtract the standard model prediction of 91.6~fb to find the upper bound on anomalous $WV\gamma$ production. We then compare to the simulated, ALP-mediated cross section to set limits on the ALP coupling.

\bibliography{alprefs}

\end{document}